\documentstyle[12pt]{article}

\input epsf

\begin{document}
\renewcommand{\thefootnote}{\fnsymbol{footnote}}
\thispagestyle{empty}
\begin{flushright}
LMU 08/96\\
hep-ph/9610543
\end{flushright}
\vspace{1cm}
\begin{center}
 \begin{Large}
{\bf Nonperturbative QCD Contributions to the Semileptonic Decay
Width of the B Meson}
 \end{Large}
\end{center}
  \vspace{1cm}
   \begin{center}
Changhao Jin\footnote{E-mail: jin@photon.hep.physik.uni-muenchen.de}\\
      \vspace{0.4cm}
    {\it Sektion Physik,  Universit\"at M\"unchen\\
        Theresienstrasse 37, D--80333 M\"unchen, Germany}
  \vspace{1cm}
\end{center}
\begin{abstract}
Nonperturbative QCD contributions to the inclusive semileptonic decay of the
$B$ meson consist of the dynamic and kinematic components.
We calculate the decay width in an approach based on the light-cone expansion
and the heavy quark effective theory, which is able to include
both components of nonperturbative QCD contributions.
The kinematic component 
results in the phase-space extension and is shown to be
quantitatively crucial, which could increase the decay width significantly. 
We find that the semileptonic decay width is enhanced by 
long-distance strong interactions by $+(9\pm 6)\%$. 
This analysis is used to determine
the CKM matrix element $\left|V_{cb}\right|$ with a controlled 
theoretical error. Implications of the phase-space effects for the
nonleptonic decay widths of $b$-hadrons are briefly discussed. 
The experimental evidence for the phase-space effects is pointed out.
\end{abstract}
\vspace{1cm}
\begin{center}
{\it published in Phys. Rev. D 56, 2928 (1997)}
\end{center}
\newpage
\section{Introduction}
A direct goal of studying the inclusive semileptonic B meson decay
$\bar B\rightarrow e\bar \nu_eX$
is to determine the standard model parameter $\left| V_{cb}\right|$
accurately. The semileptonic decay width
can be expressed as
\begin{equation}
\Gamma_{SL}=\gamma_c\left| V_{cb}\right|^2+\gamma_u\left| V_{ub}\right|^2 .
\label{eq:divd}
\end{equation}
The first term in (\ref{eq:divd}) results from the $b\rightarrow c$ 
transition. The second term in (\ref{eq:divd}) is due to 
the $b\rightarrow u$ transition and
is negligible in comparison with the first term since
$\gamma_u\sim\gamma_c$ and
$\left| V_{ub}\right|\ll \left| V_{cb}\right|$.
The semileptonic decay width is determined by two measured quantities:
the inclusive semileptonic B decay branching ratio $B_{SL}$ and
the B meson lifetime
$\tau_B$,
\begin{equation}
\Gamma_{SL}=\frac{B_{SL}}{\tau_B} .
\end{equation}
Therefore, the CKM matrix element $\left| V_{cb}\right|$ 
can be determined through
\begin{equation}
\left| V_{cb}\right|^2=\frac{\Gamma_{SL}}{\gamma_c}=
\frac{B_{SL}}{\gamma_c\tau_B} ,
\label{eq:vcb}
\end{equation}
with the theoretical input $\gamma_c$.

Theory is needed to calculate $\gamma_c$ and
to understand quantitatively
uncertainties in this calculation.
The main obstacle to this end is the difficulty of taking into account 
nonperturbative QCD effects on the underlying
weak decay process.

In recent years a heavy quark expansion approach to 
inclusive B decays has been
developed [1--7]
to account for nonperturbative QCD effects.
This approach is based on the operator product expansion and the
heavy quark effective theory (HQET).
An operator product expansion on the time ordered product of two
currents is performed. The momentum of the incoming b quark is written
as $p_b=m_bv+k$ ($m_b$ stands for the b quark mass and
$v$ the B hadron velocity) and the residual
momentum, $k$, is expanded in. For keeping track of the $m_b$ dependence
of matrix elements,
the b quark operators in full QCD are matched onto those in the HQET.
The leading term of the expansion coincides with the free quark 
decay model. The next terms are computed in powers of $1/m_b$, where
no $1/m_b$ term appears.

The calculations \cite{biur,neu2} 
in the heavy quark expansion approach
claimed that nonperturbative QCD contributions {\it decrease} the 
semileptonic decay width by a few percent with respect to the free
quark decay width.
There are, however, theoretical limitations in this approach \cite{manohar}.
The operator product expansion breaks down for low--mass final hadronic
states. In particular, the endpoint singularities of the lepton spectra
indicate a failure of the operator product expansion.
Moreover, the truncation of the expansion enforces the use of quark
kinematics rather than physical hadron kinematics.
Describing the lepton spectra demands a resummation of the heavy quark
expansion \cite{mn}.
There remains a need to clarify
the consequence of the theoretical limitations for the calculation of 
the semileptonic decay width, as it is desirable to
improve the accuracy in the independent determination\footnote{Using different
experimental and theoretical methods, 
$|V_{cb}|$ can be also determined independently from exclusive
semileptonic B decays.} 
of $|V_{cb}|$ from
the inclusive semileptonic B decay with theoretical refinements.

The resummation of the heavy quark expansion introduces \cite{mn} a distribution
function (``shape function'') of the $b$ quark in the B meson, which incorporates
nonperturbative QCD effects.
A similar distribution function arises \cite{jp,jp1} from the light--cone 
dominance
in the inclusive semileptonic B meson decays.
The introduction of the distribution function eliminates the theoretical 
difficulties mentioned above, namely the endpoint singularities are absent and
the use of physical hadron kinematics is allowed (but does not arise ``for free'').

In this paper we will use the light--cone approach \cite{jp,jp1}
to calculate the nonperturbative QCD contributions.
This approach describes the decay by using the light--cone
expansion and the HQET, 
which provides a theoretical justification for the
DIS-like parton model \cite{jin}. 
The predicted electron energy spectrum 
agrees well with the experimental measurement \cite{jp1}.
The use of physical hadron kinematics is built into this approach explicitly, 
so that
both dynamic and kinematic effects of nonperturbative QCD are properly
taken into account. The latter is shown to
be quantitatively crucial. 
We find an about $9\%$ enhancement of
the semileptonic decay width with respect to the free quark
decay width by nonperturbative QCD contributions, 
in contrast to the results obtained in the heavy quark
expansion approach. 

\begin{figure}
\centering
\epsfysize=12cm
\epsffile{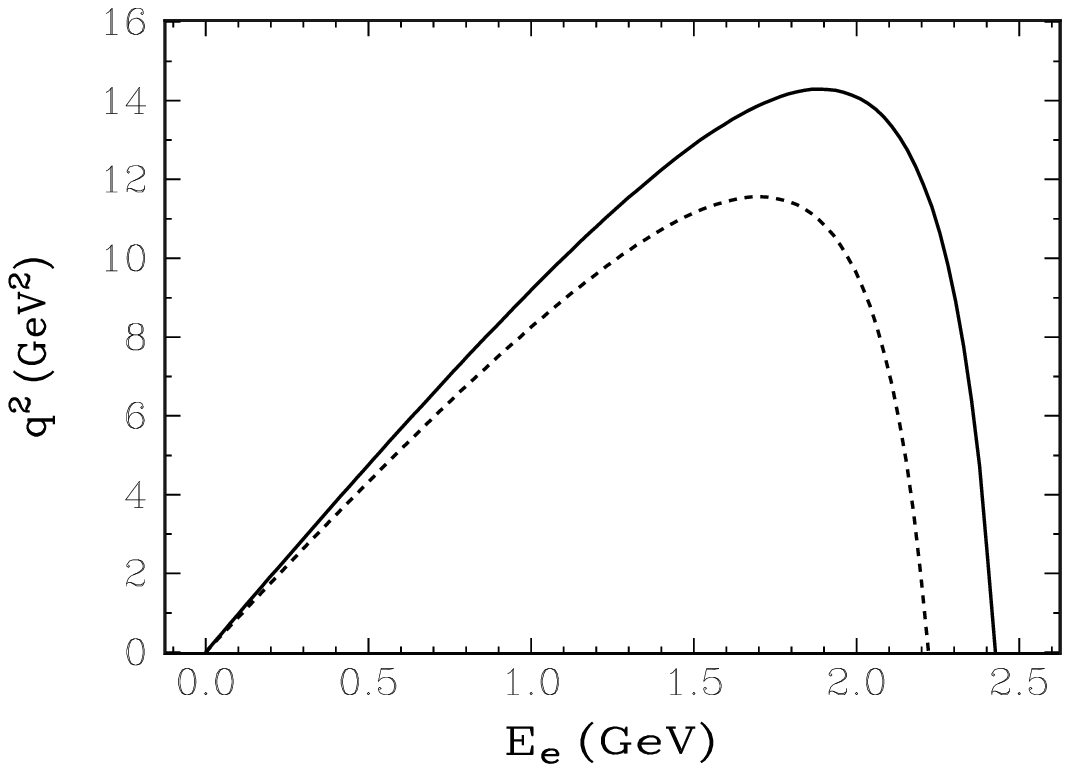}
\vspace{-0.5cm}
\caption{Phase space for the $b\rightarrow c$ inclusive semileptonic decay. 
The interior of
the solid curve is the hadron level phase space (the changeableness 
of the mass of the final hadronic state is not shown explicitly). 
The interior of the dashed
curve is the quark level phase space.} 
\label{fig:ext1}
\end{figure}

The reason of the enhancement is the following.
There are two components -- dynamics and kinematics -- of nonperturbative 
QCD effects on inclusive
semileptonic B decays. First, the decay dynamics deviates from the free
quark decay dynamics as quarks are confined in hadrons and can never be free.
However, the dynamic deviation changes the decay width only slightly since
the $b$ quark inside the B meson is almost on shell.
Second, the decay kinematics gets changed. The phase space extends
from the quark level
to the hadron level (the detailed formulas will be given below in 
(\ref{eq:ha1})--(\ref{eq:ha3}) for hadron kinematics and in
(\ref{eq:qu1}) and (\ref{eq:qu2}) for quark kinematics), 
shown in Fig.\ref{fig:ext1}
for the $b\rightarrow c$ decay.
The phase-space extension arises from the difference in the B meson
and $b$ quark masses and the fact that the mass of the decay product quark
is fixed in the free quark decay picture, while the mass of the final hadronic
state is actually changeable. The phase-space effect is a dominating factor,
as indicated by the replacement of the $b$-quark mass with the $B$-meson mass
in the decay rate.
It is thus important to include this type of contributions to
the decay width.
Consequently, it is conceivable that the net effect of nonperturbative QCD
{\it enhances} the semileptonic decay width. 
The negative contribution found in previous studies 
in the heavy quark expansion approach
is just a reflection of incompleteness of the calculation,
which fails to take into account, in particular,
a large part of nonperturbative QCD contributions due to
the phase-space effect.

We will describe the approach in section 2 and analyse  
nonperturbative QCD contributions in comparison with the heavy quark expansion
and extract then $\left|V_{cb}\right|$
from the inclusive semileptonic B meson decay in section 3 and 
finally conclude and discuss in section 4.

\section{Approach}
The semileptonic decay width can be split into two parts:
one, denoted by $\Gamma_{nonpert}$, includes nonperturbative QCD
contributions,
the other
results from perturbative QCD corrections to the decay width, denoted by
$\Gamma_{pert}$. Namely,
\begin{equation}
\Gamma_{SL}=\Gamma_{nonpert}+\Gamma_{pert}.
\label{eq:sum}
\end{equation}

Nonperturbative QCD effects are contained in the hadronic tensor
$W_{\mu\nu}$. It can be written in terms of a current commutator
taken between B states:
\begin{equation}
W_{\mu\nu}= -\frac{1}{2\pi}\int d^4ye^{iq\cdot y}
\langle B\left|[j_\mu (y),j_\nu^\dagger (0)]\right|B\rangle ,
\end{equation}
where $q$ is the momentum transfer to the final lepton pair.
$\mid \mkern -6mu B \rangle$ refers to the B-meson state with energy $E_B$ 
and is normalized according to 
$\langle B\mkern -6mu\mid \mkern -6mu B\rangle = 
2E_B(2\pi)^3\delta^3({\bf 0})$.

It is well known that integrals like the one in eq.(5) are dominated by
distances where
\begin{equation}
0\leq y^2\leq\frac{1}{q^2} .
\end{equation}
For inclusive semileptonic B-meson decays, the momentum transfer squared 
$q^2$ is timelike and varies in the physical range
\begin{equation}
0\leq q^2\leq (M_B-M_{X_{min}})^2,
\end{equation}
where $M_B$ and $M_{X_{min}}$ represent the B-meson mass and the minimum value of 
the invariant mass of the hadronic final state, respectively. 
Due to the large B-meson mass, extended regions of
phase space involve large values of $q^2$ (see also Fig.1). 
Therefore, the decay is dominated
by light--cone distances between the two currents in eq.(5).  
This allows to replace the
commutator of the two currents with its singularity on the light cone
times an operator bilocal in the b quark fields.
Furthermore, the light-cone dominance enables us to expand the matrix element
of the bilocal operator between B-meson states in powers of
$\Lambda_{QCD}^2/q^2$. The leading nonperturbative effect is described
by a distribution function \cite{jp,jp1}:
\begin{equation}
f(\xi)=\frac{1}{4\pi M_B^2}\int d(y\cdot P_B)e^{i\xi y\cdot P_B}
\langle B\left| \bar b(0)/ \mkern -12mu P_B(1-\gamma_5)b(y)
\right|B\rangle \
|_{y^2=0} ,
\label{eq:dlight}
\end{equation}
where $P_B$ denotes the four-momentum of the B meson.
$f(\xi)$ is the probability of finding a $b$-quark with momentum $\xi P_B$
inside the B meson. The hadronic tensor can be expressed in terms of
the distribution function: 
\begin{equation}
W_{\mu\nu}= 4(S_{\mu\alpha\nu\beta}-i\varepsilon_{\mu\alpha\nu\beta})
\int d\xi f(\xi)\varepsilon (\xi P_{B_0}-q_0)\delta [(\xi P_B-q)^2-m_c^2]
(\xi P_B-q)^\alpha P_B^\beta ,
\end{equation}
where $m_c$ is the charm quark mass.

$\Gamma_{nonpert}$ is calculated in this approach 
by integrating the differential decay rate
in the B rest frame,
\begin{equation}
\Gamma_{nonpert}=\int dE_e\int dq^2\int dM_X^2
\frac{d^3\Gamma}{dE_edq^2dM_X^2} ,
\label{eq:nonpert}
\end{equation}
with the differential decay rate
\begin{equation}
\frac{d^3\Gamma}{dE_edq^2dM_X^2}=
\frac{G_F^2\left|V_{cb}\right|^2}{8\pi^3M_B}
\frac{q_0-E_e}{\sqrt{{\bf q}^2+m_c^2}} 
\Bigg \{f(\xi_+)
\Bigg (2E_e\xi_+-\frac{q^2}{M_B}\Bigg ) 
-(\xi_+ \rightarrow \xi_-)\Bigg \} ,
\label{eq:nonpert1}
\end{equation}
where $E_e$ is the electron energy and we have
neglected the electron mass. 
$M_X$ denotes the invariant mass of the final hadronic state.
The dimensionless variables $\xi_\pm$ reads
\begin{equation}
\xi_\pm=\frac{q_0\pm\sqrt{{\bf q}^2+m_c^2}}{M_B}.
\end{equation}
Note that there appears in the differential decay rate (\ref{eq:nonpert1}) 
the B meson mass rather than the b quark mass.
The integration limits are specified by hadron kinematics:
\begin{eqnarray}
0\leq&E_e&\leq \frac{M_B}{2}\Bigg (1-
\frac{M^2_{X_{min}}}{M_B^2}\Bigg ), \label{eq:ha1} \\
0\leq&q^2&\leq 2E_e\Bigg (M_B-
\frac{M^2_{X_{min}}}{M_B-2E_e}\Bigg ), \label{eq:ha2} \\
M^2_{X_{min}}\leq&M_X^2&\leq (M_B-2E_e)
\Bigg (M_B-\frac{q^2}{2E_e}\Bigg ) , \label{eq:ha3}
\end{eqnarray}
which define the hadron level phase space shown in Fig.1.
For comparison, we also write down here the kinematic boundaries for the free
quark decay:
\begin{eqnarray}
0\leq&E_e&\leq \frac{m_b}{2}\Bigg (1-\frac{m_c^2}{m_b^2}\Bigg ), \label{eq:qu1}\\
0\leq&q^2&\leq 2E_e\Bigg (m_b-\frac{m_c^2}{m_b-2E_e}\Bigg ), \label{eq:qu2}
\end{eqnarray}
which are also shown in Fig.1.
It is an important feature of this approach that the calculation can be
performed in the physical phase space as a large contribution of
nonperturbative QCD arises from the extension of phase space from
the quark level to the hadron level. 
It should be pointed out, however, that in theoretical calculations we
take $M_{X_{min}}=m_c$ since we assume quark-hadron duality in our
approach.

Important properties of the distribution function are derived from
field theory. Due to current conservation, it is exactly normalized to unity 
with a support $0\leq\xi\leq 1$.
It obeys positivity. When the distribution function becomes the delta
function, $\delta(\xi-m_b/M_B)$, the free quark decay is reproduced.
Furthermore, the next two moments of the distribution function
can be estimated in the HQET, as we shall discuss below.
These two moments determine the mean value $\mu$ and the variance
$\sigma^2$ of the distribution function,
which characterize the position of the maximum and the width of 
it, respectively:
\begin{eqnarray}
& &\mu\equiv M_1(0)=\tilde{\xi}+M_1(\tilde{\xi}) ,
\label{eq:pmean} \\
& &\sigma^2\equiv M_2(\mu)=M_2(\tilde{\xi})-M_1^2(\tilde{\xi}) ,
\label{eq:pvar}
\end{eqnarray}
where $M_n(\tilde{\xi})$ is
the nth moment about a point $\tilde{\xi}$ of the  
distribution function defined by
\begin{equation}
M_n(\tilde{\xi})=\int_0^1 d\xi  (\xi-\tilde{\xi})^nf(\xi) .
\end{equation}
By definition, $M_0(\tilde{\xi})=1$.

The accuracy of the theory is remarkably improved 
by estimating the next two
moments of the distribution function in the framework of the HQET.
However, it cannot yet be completely determined in QCD.
For practical calculations, therefore, we shall use an ansatz for 
the distribution
function, which respects all known properties, with two parameters $a$
and $b$ as follows
\begin{equation}
f(\xi)=N\frac{\xi (1-\xi)}{(\xi-a)^2+b^2}\theta(\xi)\theta(1-\xi) \ ,
\label{eq:ansatz}
\end{equation}
where N is the normalization constant.
For $a=m_b/M_B$ and $b=0$, eq.(\ref{eq:ansatz}) becomes a delta function,
$\delta(\xi-m_b/M_B)$, and the free quark decay is reproduced.
Another form of the distribution function has been proposed in \cite{LK}.

The perturbative QCD corrections to $O(\alpha_s)$ has been 
calculated \cite{cabi,nir}.
It has the form
\begin{equation}
\Gamma_{pert}= -\frac{2\alpha_s}{3\pi}H\Bigg (\frac{m_c}{m_b}\Bigg )
\Gamma_b,
\label{eq:pert}
\end{equation}
where $\Gamma_b$ is the free quark decay width:
\begin{equation}
\Gamma_b=\Gamma_0\Phi\Bigg (\frac{m_c}{m_b}\Bigg ) ,
\end{equation}
with 
\begin{equation}
\Gamma_0=\frac{G_F^2m_b^5\left|V_{cb}\right|^2}{192\pi^3},
\end{equation}
\begin{equation}
\Phi(x)=1-8x^2+8x^6-x^8-24x^4lnx.
\end{equation}
An analytic expression for $H(m_c/m_b)$ is given in \cite{nir}.

The semileptonic decay width can be calculated by substituting 
(\ref{eq:nonpert}) and (\ref{eq:pert}) into (\ref{eq:sum}).
The parameters involved and hence the sources of the theoretical error are:\\
(1) the parameters in the distribution function (for the ansatz
(\ref{eq:ansatz}) that will be used, they are $a$ and $b$),  \\
(2) the beauty and charm quark pole masses $m_b$ and $m_c$, \\
(3) the strong coupling constant $\alpha_s$.

There are several theoretical constraints on these parameters stemming from
the HQET, which reduce the theoretical uncertainties considerably.  
We discuss them in turn. \\
{\bf I}.  Performing a light-cone OPE and following the method 
of \cite{manohar}
to expand the matrix elements of the local operators in the HQET, 
the next two moments and hence the mean value
$\mu$ and the variance $\sigma^2$ of the distribution function
can be related \cite{jp} to two accessible parameters $K_b$ and $G_b$
up to the order of $(\Lambda_{QCD}/m_b)^2$,
\begin{eqnarray}
&& \mu=\frac{m_b}{M_B}(1+E_b) , \\
&& \sigma^2=\Bigg (\frac{m_b}{M_B}\Bigg )^2\Bigg (\frac{2K_b}{3}-E_b^2\Bigg ) ,
\end{eqnarray}
where $E_b=K_b+G_b$. Both $K_b$ and $G_b$ are of order 
$(\Lambda_{QCD}/m_b)^2$. This leads to a model-independent
conclusion: the distribution function is sharply peaked around
$\xi=\mu\approx m_b/M_B$ and its width is of order $\Lambda_{QCD}/M_B$.

For numerical analyses we need to know $K_b$ and $G_b$ quantitatively.
The parameter $G_b$ is related \cite{manohar} to the observables,
\begin{equation}
m_bG_b=-\frac{3}{4}(M_{B^\ast}-M_B),
\end{equation}
where the mass difference of the vector $B^\ast$ and the pseudoscalar $B$
mesons is measured to be $M_{B^\ast}-M_B=0.046$ GeV.
$K_b$ can be reexpressed in terms of another
often used parameter $\lambda_1$ instead,
\begin{equation}
K_b=-\frac{\lambda_1}{2m_b^2} .
\end{equation}
It is harder to determine $\lambda_1$ (or $K_b$).
The accurate value of it is not known.
Consequently, the mean value $\mu$ and the variance $\sigma^2$ of the
distribution function are determined by the two parameters $m_b$ and
$\lambda_1$: $\mu$ depends on $m_b$ strongly and $\lambda_1$ very
weakly, while $\sigma^2$ is sensitive essentially only to $\lambda_1$.
Hence, the parameters $a$ and $b$ in the ansatz (\ref{eq:ansatz}) for
the distribution function are also determined by $m_b$ and $\lambda_1$. \\
{\bf II}. The quark mass difference is related to $\lambda_1$
in the HQET \cite{neu2}
\begin{equation}
m_b-m_c=(\overline{M}_B-\overline{M}_D)\Bigg\{1-\frac{\lambda_1}
{2\overline{M}_B\overline{M}_D}
+{\cal O}(1/m_c^3)\Bigg\},
\end{equation}
where the spin-averaged meson masses
\begin{eqnarray}
\overline{M}_B & = & \frac{1}{4}(M_B+3M_{B^\ast})=5.31 \ GeV , \\
\overline{M}_D & = & \frac{1}{4}(M_D+3M_{D^\ast})=1.97 \ GeV.
\end{eqnarray}

Finally, the remaining theoretical input parameters for our analysis
are $m_b$, $\lambda_1$,
and $\alpha_s$. 

\section{Analysis}
We evaluate $\Gamma_{nonpert}$ and $\Gamma_{pert}$ in the approach described
above using the three input parameters. 
For $m_b$ we use
\begin{equation}
m_b=4.9\pm 0.2 \ \mbox{GeV}.
\end{equation}
According to a QCD sum rule calculation \cite{ball},
we take
\begin{equation}
\lambda_1=  -(0.5\pm 0.2) \ \mbox{GeV}^2.
\end{equation}
As a result, the mean value and the variance of the distribution function
are:
\begin{eqnarray}
\mu&=&0.93\pm 0.04 , \\
\sigma^2&=&0.006\pm 0.002.
\end{eqnarray}
A truncating of perturbative series causes the dependence of 
perturbative calculations on
the renormalization scale $\mu_r$.
For inclusive semileptonic B decays perturbative QCD corrections are
known only to the leading order. The result given in (\ref{eq:pert})
exhibits an implicit scale dependence of the strong coupling $\alpha_s$.
We vary the scale over the range of $m_b/2\leq\mu_r\leq m_b$ to estimate
the theoretical error due to the choice of the scale used in the argument
of $\alpha_s$.

The dependence of the decay width $\Gamma$ on the parameters is 
shown in Fig.2.
The variation of $\Gamma$ with $m_b$ or $\lambda_1$ is stronger than $\mu_r$.
The variation of $m_b$ leads to an uncertainty of $8\%$ in the decay width
if other parameters are kept fixed. The same uncertainty in the decay width
results from the variation of $\lambda_1$. An uncertainty of $2\%$ 
in the decay width is
introduced when the renormalization scale $\mu_r$ is varied between $m_b/2$
and $m_b$. In addition, the impact of the shape of the distribution function
on the value of the decay width is studied. The value of the decay width
is more sensitive to the variation of the mean value than 
the variation of the variance of
the distribution function. 
Furthermore, we modify (\ref{eq:ansatz}) 
with two more parameters $\alpha$ and $\beta$ to be
\begin{equation}
f(\xi)=N\frac{\xi (1-\xi)^\alpha}{[(\xi-a)^2+b^2]^\beta}\theta(\xi)
\theta(1-\xi) \ .
\end{equation}
Using (37) we find that the value of the decay width is insensitive to 
the change of 
the shape of the distribution function if the mean value and 
the variance of it are kept fixed.
This insensitivity diminishes the model dependence.  
This analysis yields
\begin{equation}
\gamma_c= 49\pm 9 \ \mbox{ps$^{-1}$},
\label{eq:gammac}
\end{equation}
with a theoretical error of $18\%$.

In Fig.3, we compare the decay widths calculated in our approach, the free
quark decay model, and the heavy quark expansion approach.
The result in our approach shows that nonperturbative QCD contributions 
enhance the decay width by $+(9\pm 6)\%$ with respect to the free quark
decay width, in contrast to the result of the heavy quark expansion approach
where a reduction of the free quark decay width by $-(4.3\pm 0.5)\%$
is found.
The change of the sign indicates that the nonperturbative effects receive 
a large phase-space enhancement. We also observe that the decay width 
calculated in our approach
goes to a free quark decay limit as $-\lambda_1$ decreases.
This behavior is expected since the distribution function approaches
a delta function, which reproduces the free quark decay, 
as $-\lambda_1$ and hence $\sigma^2$ decrease.
Thus this behavior provides a check of calculations. 
   
This theoretical analysis can be used to determine $\left|V_{cb}\right|$.
Experimentally
the inclusive semileptonic branching ratio $B_{SL}$ 
has been measured at the
$\Upsilon(4S)$ and $Z^0$ resonances, respectively.
The lifetime $\tau_B$ has been 
measured by experiments at $Z^0$ and in $p\bar p$ collisions.
The average of these measurements leads to \cite{skw}
\begin{equation}
\Gamma_{SL}=67.3\pm 2.7 \ \mbox{ns$^{-1}$}.
\label{eq:width}
\end{equation}
Putting it together with the theoretical value of $\gamma_c$
given in (\ref{eq:gammac}), we obtain from (\ref{eq:vcb})
\begin{equation}
\left| V_{cb}\right|=0.0371\pm 0.0007\pm 0.0034 ,
\end{equation}
where the first error is experimental and the second theoretical.

\section{Conclusion and Discussion}
We have calculated the semileptonic decay width of the B meson using an 
approach based on the light-cone expansion and the HQET.
Nonperturbative QCD effects are described by a single distribution function.
Several important properties of the distribution function are known from
QCD and the HQET of it.
However, one still has to model the distribution function.
Fortunately the result of the calculation of the decay width in this approach
is nearly model-independent, since it is essentially only sensitive to
the mean value and the variance of the distribution function, whose 
theoretical estimates exist.
Moreover, this approach is able to take into account both dynamic
and kinematic components of nonperturbative QCD effects. We have shown that 
including the latter is indeed quantitatively crucial, 
which could increase the decay width significantly. 
We find an enhancement of the free quark decay width by $+(9\pm 6)\%$
due to nonperturbative QCD contributions, contrary to the claims from
the heavy quark expansion approach.
As a result, a value of $\left|V_{cb}\right|$ is extracted from the inclusive
semileptonic B meson decay with a controlled theoretical error.

The main theoretical uncertainty arises from the values of the b quark mass
and the HQET parameter $\lambda_1$.
It seems possible to reduce theoretical uncertainties 
by a detailed fit\footnote{Such a fit has been done in the heavy quark
expansion approach \cite{gremm}.} 
to the measured charged-lepton energy spectrum to determine 
the parameters 
and a calculation of the 
next-to-leading order perturbative QCD correction.\footnote{Partial 
calculations of higher-order corrections exist \cite{high}.}
Future measurements
of the distribution function and more theoretical efforts on calculations
of hadronic matrix elements should enable to further reduce the uncertainties.

Careful inclusion of the kinematic effect of nonperturbative strong 
interactions is also necessary for reliable predictions for the nonleptonic
decay widths of hadrons containing a $b$ quark. 
The nonleptonic decay widths of $b$ hadrons 
may be calculated in a similar way. 
The decay widths
are expressed in terms of the $b$ hadron masses rather than the $b$ quark mass
provided the phase-space effects are included, whereas according to
the heavy quark expansion, the relevant mass in the decay widths should be 
the universal $b$-quark mass and no corrections of order $1/m_b$ should
be present \cite{non}. 
We would anticipate
an enhancement of the nonleptonic decay width by nonperturbative
QCD if both dynamic and kinematic effects of it are properly taken 
into account. 
Since the phase-space effects cancel out to a large extent in the
ratio of the decay widths, they are unlikely to significantly change
the semileptonic branching ratio of the $B$ meson or the average number of 
charmed hadrons produced per B decay. On the other hand, the prediction
on the ratio of the $\Lambda_b$ and $B$ lifetimes from the heavy quark 
expansion approach seems to be in conflict with the data \cite{rew}. 
In this case phase-space effects are enlarged due to the significant difference 
between the $\Lambda_b$ and $B$ masses.
The replacement of the $b$-quark mass with the non-universal $b$-hadron masses 
results in a perfect agreement \cite{am} between the theory and 
the experimental data, giving evidence for the phase-space effects. 
\vspace{1cm}
\begin{flushleft}
{\bf Acknowledgements}
\end{flushleft}
I would like to thank E.A. Paschos for discussions.

\newpage

\newpage
\begin{figure}
\centering
\epsfxsize=15cm
\epsffile{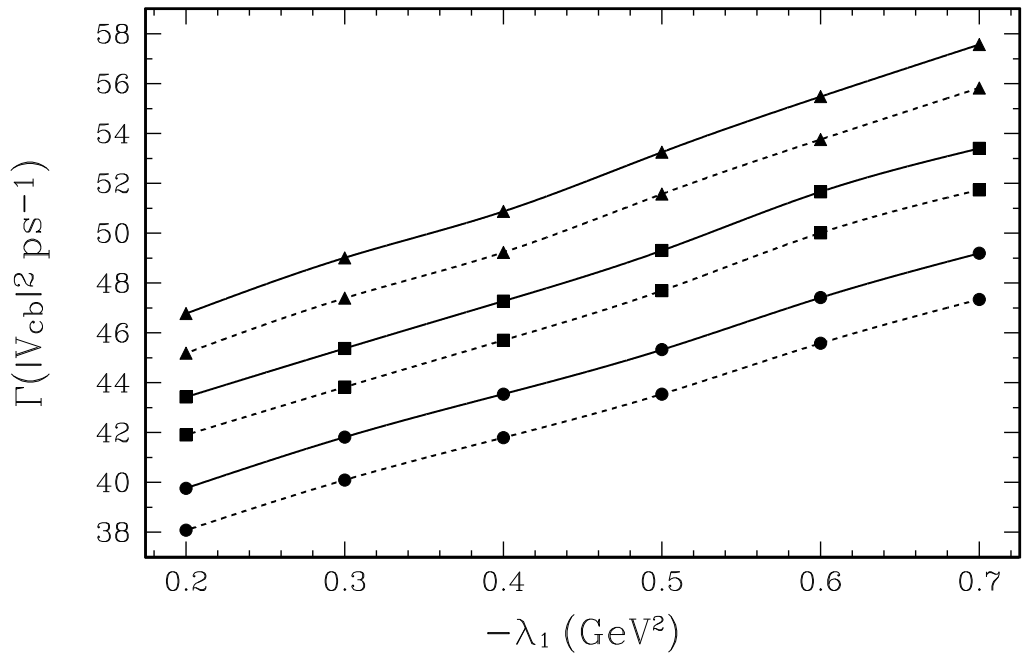}
\vspace{-0.5cm}
\caption{Dependence of the semileptonic decay width $\Gamma$ on the
theoretical input parameters $m_b$, $\lambda_1$, and $\mu_r$.
The solid (dashed) curves are for the renormalization scale $\mu_r=m_b$
($\mu_r=m_b/2$). The curves with solid dots, boxes, triangles correspond to
$m_b=4.7, \ 4.9, \ 5.1$ GeV, respectively.}
\label{fig:width}
\end{figure}

\newpage
\begin{figure}
\centering
\epsfxsize=15cm
\epsffile{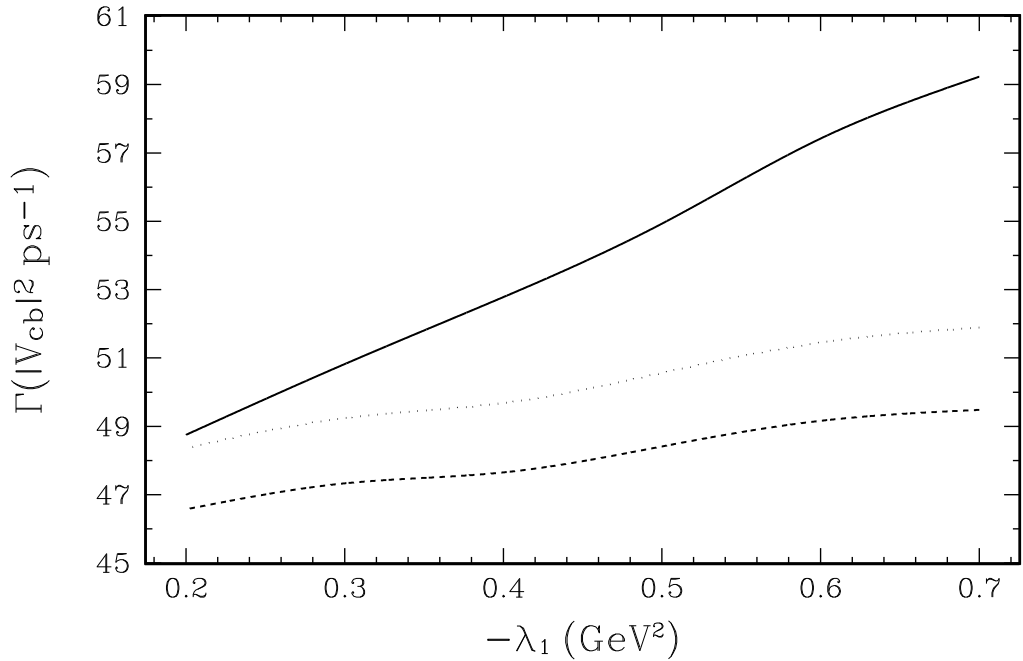}
\vspace{-0.5cm}
\caption{Semileptonic decay width $\Gamma$ as a function of $\lambda_1$
calculated in our approach (solid curve), the free quark decay model
(dotted curve), and the heavy quark expansion approach (dashed curve).
We take $m_b=4.9$ GeV and $\alpha_s=0$.}
\label{fig:com}
\end{figure}

\end{document}